\documentclass[aps,prl,twocolumn,superscriptaddress,floatfix]{revtex4}

\usepackage{graphicx}
\usepackage{color}
\usepackage{amsmath}
\usepackage[babel]{csquotes}
\usepackage{slashed}

\begin{document}

\title{Access to improve the muon mass and magnetic moment anomaly via the bound-muon $g$ factor}

\author{B. Sikora}
\email[]{bsikora@mpi-hd.mpg.de}
\affiliation{Max Planck Institute for Nuclear Physics, Saupfercheckweg 1, 69117 Heidelberg, Germany}
\author{H. Cakir}
\affiliation{Max Planck Institute for Nuclear Physics, Saupfercheckweg 1, 69117 Heidelberg, Germany}
\author{N. Michel}
\affiliation{Max Planck Institute for Nuclear Physics, Saupfercheckweg 1, 69117 Heidelberg, Germany}
\author{V. Debierre}
\affiliation{Max Planck Institute for Nuclear Physics, Saupfercheckweg 1, 69117 Heidelberg, Germany}
\author{N.~S.~Oreshkina}
\affiliation{Max Planck Institute for Nuclear Physics, Saupfercheckweg 1, 69117 Heidelberg, Germany}
\author{N.~A. Belov}
\affiliation{Max Planck Institute for Nuclear Physics, Saupfercheckweg 1, 69117 Heidelberg, Germany}
\author{V.~A. Yerokhin}
\affiliation{Max Planck Institute for Nuclear Physics, Saupfercheckweg 1, 69117 Heidelberg, Germany}
\affiliation{Center for Advanced Studies, Peter the Great St.~Petersburg Polytechnic University, 195251 St.~Petersburg, Russia}
\author{C.~H. Keitel}
\affiliation{Max Planck Institute for Nuclear Physics, Saupfercheckweg 1, 69117 Heidelberg, Germany}
\author{Z. Harman}
\email[]{harman@mpi-hd.mpg.de}
\affiliation{Max Planck Institute for Nuclear Physics, Saupfercheckweg 1, 69117 Heidelberg, Germany}

\date{\today}

\begin{abstract}

A theoretical description of the $g$ factor of a muon bound in a nuclear potential is presented. One-loop self-energy and multi-loop vacuum
polarization corrections are calculated, taking into account the interaction with the binding potential exactly. Nuclear effects on the bound-muon
$g$ factor are also evaluated.
We put forward the measurement of the bound-muon $g$ factor via the continuous Stern-Gerlach effect as an independent means to determine the free muons
magnetic moment anomaly and mass. The scheme presented enables to increase the accuracy of the mass by more than an order of magnitude.

\end{abstract}

\keywords{$g$ factor,muonic atom}

\maketitle

The physics of muons features puzzling discrepancies. The disagreement of the free muons experimental and theoretical $g$ factor by $3\sigma$
represents the largest deviation from the Standard Model observed in an electroweak quantity~\cite{Jegerlehner2009}. Recently, high-precision
spectroscopy experiments with the muonic H atom yielded a value for the proton radius which strongly disagrees with that obtained from
measurements on regular H~\cite{Pohl2010,Antognini2013} (see also~\cite{Beyer2017}). Therefore, experiments aiming at improved
determinations of the muons properties help to clarify these issues and can be a hint for New Physics.

The fast progress in the theoretical understanding and experimental precision of the bound-electron $g$ factor (see e.g.
\cite{Haffner2000,Verdu2004,Sturm2011,Sturm2011b,BeierReport,Yerokhin2004,Pachucki2004,Pachucki2017,Czarnecki2017} and references therein)
has also enabled the most accurate determination of the mass of the \textit{electron} in Penning trap $g$-factor experiments by means of
the continuous Stern-Gerlach effect~\cite{Haffner2000,Beier2001,Sturm2014,Kohler15,Zatorski2017}. In this Letter we put forward a similar
method for the extraction of the mass of the \textit{muon} by employing light muonic ions, by which we mean here bound systems solely
consisting of a nucleus and a muon without further surrounding electrons. Since currently the mass of the muon is only known from
muonium spectroscopy~\cite{Jungmann2000} and to a fractional standard uncertainty of 2.2$\times 10^{-8}$~\cite{CODATA2014,CODATA2014b}, alternative
methods for its determination are especially desirable.

When a muonic ion is subjected to a magnetic field of strength $B$,
the Larmor frequency between the bound-muon Zeeman sublevels depends on the magnetic moment $\mu$ of the muon by the formula
\begin{equation}
\label{eq:Larmor}
\omega_{\rm L} = \frac{ 2 \mu}{\hbar} B = \frac{g}{2} \frac{e}{m_{\mu}} B\,,
\end{equation}
with $e$ being the (positive) unit charge, and $g$ and $m_{\mu}$ the bound muons $g$ factor and mass, respectively. Determining
the magnetic field at the location of the ion becomes possible through a measurement of the cyclotron frequency of the ion,
\begin{equation}
\label{eq:cycl}
\omega_{\rm c} = \frac{Q}{M} B \,,
\end{equation}
where $Q$ and $M$ are the charge and mass of the muonic ion, respectively. Thus, $m_{\mu}$ can be expressed by $M$ as
\begin{equation}
\label{eq:massdet}
m_{\mu} = \frac{g}{2} \frac{e}{Q} \frac{\omega_{\rm c}}{\omega_{\rm L}} M\,,
\end{equation}
where the theoretical value $g_{\rm theo}$ for the bound-muon $g$ factor is to be substituted. The quantity to be measured is the
ratio of the two frequencies, $\Gamma={\omega_{\rm L}}/{\omega_{\rm c}}$. For determining $m_{\mu}$ with a given fractional uncertainty,
all the quantities $g_{\rm theo}$, $\Gamma$ and $M$ have to be known at the same level of accuracy. Alternatively, Eq.~(\ref{eq:Larmor})
and (\ref{eq:cycl}) can be combined to yield an experimental bound-muon $g$ factor
\begin{equation}
\label{eq:gdet}
g = 2\frac{m_{\mu}}{M} \frac{Q}{e} \Gamma\,.
\end{equation}
Such a determination of $g = 2 + 2a_{\mu} + \Delta g_{\rm bind}$ constitutes an alternative access to the free muons magnetic moment
anomaly $a_{\mu}$ at a level at which $m_{\mu}$ is known from an independent experiment, and provided the binding contribution $\Delta g_{\rm bind}$
can be calculated to sufficient accuracy.

\begin{figure}
\includegraphics[width=0.9\columnwidth]{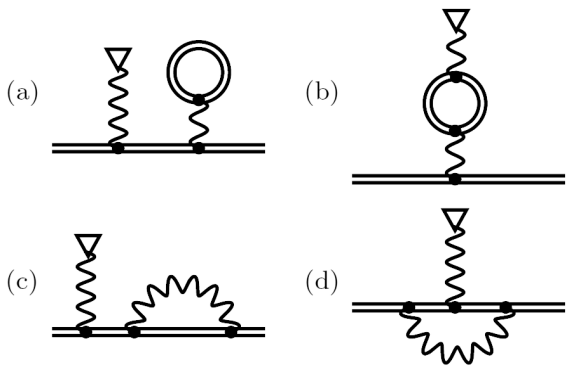}
\caption{\label{fig:QED1loop} Furry-picture Feynman diagrams depicting the one-loop QED corrections to the bound-muon
$g$ factor: (a) the electric and (b) magnetic loop vacuum polarization corrections, and (c) the self-energy wave function and (d) self-energy
vertex correction terms. A double external line represents a muonic Coulomb-Dirac wave function, and the wave line terminated by a triangle
stands for the interaction with the external magnetic field. The internal wave line represents a photon propagator, and the internal
double line depicts a Coulomb-Dirac muon propagator; in vacuum polarization loops, it may also represent an
electron-positron propagator.}
\end{figure}

In the electron mass experiments~\cite{Sturm2014,Kohler15,Zatorski2017,Haffner2000}, $^{12}$C$^{5+}$ ions were employed because the
atomic mass unit is defined in terms of the mass of the $^{12}$C atom. In determining the muon mass in a similar fashion, a lighter element,
namely $^{4}$He is more appropriate to minimize uncertainties due to nuclear effects. In addition, the mass of $^{4}$He is known to sufficient accuracy.
Therefore, in the following, we present the theory of the $g$ factor of muonic $^4$He$^{+}$ and show that a 9-digit fractional accuracy
is achievable, which corresponds to the same accuracy in the extracted muon mass or magnetic moment anomaly, provided the ratio of the Larmor and
cyclotron frequencies can be measured with matching precision.

\textit{Theoretical approach --}
The Dirac value $g_{\rm D}$ corresponds to the leading tree-level Feynman diagram with the assumption of a pointlike nucleus. It was
first calculated by Breit in 1928 \cite{Breit1928}. For a Dirac particle in the $1s$ state of an ion with a charge number $Z$ it is
$g_{\rm D} = \frac{2}{3} + \frac{4}{3} \sqrt{1 - (Z \alpha)^2}$, where $\alpha$ is the fine-structure constant. Various effects
shift the bound-muon $g$ factor from this value: Firstly, due to the finite size of the nucleus, the interaction potential between
the muon and the nucleus deviates from a pure Coulomb potential on the fm scale. Therefore, the wave function of the bound muon and
hence its $g$ factor deviate from the corresponding quantities computed for a pure Coulomb potential. This finite size (FS) correction
to the bound-muon $g$ factor can be expressed with the nuclear root-mean-square radius $r_{\rm RMS}$ by the approximate formula~\cite{Karshenboim2000}
$\Delta g_{\rm FS} = \frac{8 m_{\mu }^2}{3} (Z \alpha)^4 r_{\rm RMS}^2 + \mathcal{O} \left( (Z \alpha)^6 \right)$,
in agreement with~\cite{Zatorski2012}. As one can see on this formula, the FS correction for bound muons is more than 4 orders of
magnitude larger than for bound electrons. The accuracy of $\Delta g_{\rm FS}$ is mostly limited by the uncertainty of $r_{\rm RMS}$.
The correction due to the deformation of the nuclear charge distribution was estimated using the method
described in \cite{Zatorski2012} and nuclear data from \cite{Frosch1967}, and was taken to be less than $10^{-14}$. We also assume a
negligibly small magnitude for the nuclear polarization correction \cite{Nefiodov2002,Volotka2014}.

The leading quantum electrodynamic (QED) corrections correspond to the one-loop Feynman diagrams shown on Fig.~\ref{fig:QED1loop}.
These diagrams represent the electric and magnetic loop vacuum polarization (VP) corrections [Fig.~\ref{fig:QED1loop}~(a) and (b), respectively]
and the self-energy (SE) wave function and vertex corrections [Fig.~\ref{fig:QED1loop}~(c) and (d), respectively]. As in free-particle QED,
these loop diagrams are ultraviolet (UV) divergent. The renormalization procedure used to cancel the divergences is based on
the expansion of the internal fermion lines in each diagram in powers of interactions with the nuclear potential.
We apply the two-time Green's function method~\cite{Shabaev2002Report} for obtaining expressions for the individual terms.

The fermion loop in the \textit{VP electric loop} diagram modifies the nuclear potential at distances on the scale of the Compton
wavelength of the loop particle. In a good approximation, this can be reduced to a free fermion loop with one interaction with the nuclear
field, leading to the Uehling potential $V_{\rm Ueh} (r)$ \cite{Uehling1935}. The effect of the Uehling term was evaluated in different ways.
First, the $g$ factor contribution of the first-order Uehling diagram can be calculated as
\begin{equation}
\Delta g_{\rm Ueh} = - \frac{8 m_\mu}{3} \langle a \vert V_{\rm Ueh} \vert \delta a \rangle\,,
\end{equation}
where $\vert a \rangle$ is the bound-muon Dirac wave function and $\vert \delta a \rangle$ is the wave function linearly perturbed by the
magnetic interaction. For a point-like nucleus, $\vert \delta a \rangle$ is known analytically~\cite{Shabaev2003}. Since the Uehling potential
does not depend on the mass of the bound particle, but only on the mass of the particle in the loop, the Uehling term can also be computed as
\begin{equation}
\Delta g_{\rm Ueh} = -  \frac{4}{3 m_\mu} \langle a \vert \frac{\partial V_{\rm Ueh}}{\partial r} \vert a \rangle\,,
\end{equation}
according to the method described in \cite{Karshenboim2005}. In both cases, the $g$-factor contribution was obtained by numerical integration,
yielding an excellent numerical agreement between the two methods. In the pointlike nuclear model, the results were also compared to the exact
analytical formula~\cite{Karshenboim2001}.
We note that $Z\alpha$ expansion results derived for electronic atoms can not be straightforwardly applied to the case of mounic atoms,
since they assume the loop particle to be identical to the bound particle.
Furthermore, electronic VP effects would be largely overestimated by $Z \alpha$ expansion formulas, thus they need to be calculated to all
orders in this parameter even at low $Z$.

The higher-order term of the electric loop VP diagram, the Wichmann-Kroll contribution, was calculated with the method of
Ref.~\cite{Fainshtein1991} and was found to be negligible. Hadronic VP corrections were estimated from the muonic Uehling term,
following Ref.~\cite{Friar1999}, as $\Delta g_{\rm VP}^{\rm had}=0.671(15)\Delta g_{\rm Ueh}^{\mu}$.

\begin{figure}
\includegraphics[width=0.8\columnwidth]{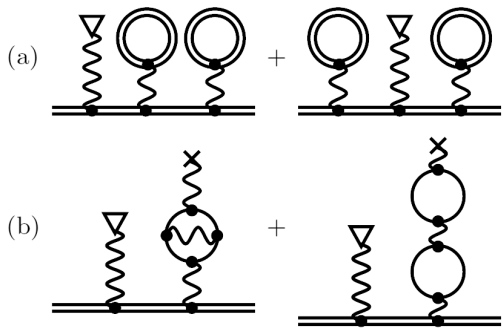}
\caption{\label{fig:QED2loop} Feynman diagrams of two-loop VP corrections to the bound-muon
$g$ factor: (a) 2nd-order electric-loop VP (Uehling and Wichmann-Kroll) terms, and (b) the K\"all\'en-Sabry diagrams. A single
internal line represents a free Dirac propagator, and the wave line terminated by a cross stands for the interaction with the
nuclear potential.}
\end{figure}

The contribution of the Uehling potential was also evaluated in an all-order treatment by including it in the radial Dirac equation,
and calculating the bound-muon wave function numerically in a B-spline representation~\cite{Shabaev2004}, as described in
Ref.~\cite{Michel2017}. This allows the extraction of the 2nd-order Uehling corrections, shown in Fig.~\ref{fig:QED2loop}~(a).
Finally, the K\"all\'en-Sabry two-loop VP correction~\cite{Kaellen1955}, illustrated in Fig.~\ref{fig:QED2loop}~(b), was evaluated
employing B-splines, and the effective potential given in Ref.~\cite{Fullerton1976,Indelicato2013}.

The lowest-order term in the expansion of the \textit{magnetic loop} VP diagram [Fig.~\ref{fig:QED1loop}~(b)] corresponds to the diagram
with the Coulomb-Dirac propagator replaced by the free Dirac propagator. This diagram is UV divergent, and its $g$-factor contribution
is canceled by charge renormalization~\cite{BeierReport}. Higher-order contributions to this diagram, such as the virtual light-by-light
scattering (LBL) term, are finite. We evaluate the LBL term as it was performed in Ref.~\cite{Lee2005,Lee2007}, with the difference
that we include the finite nuclear size effect in the bound muon wave function. Also, we calculate the mixed magnetic and electron
loop effect by repeating the above calculation with the inclusion of the effect of the Uehling potential in the bound-muon wave
function. The corresponding two-loop contribution is slightly below the uncertainty at which we aim. We note that further two-loop
VP corrections evaluated very recently for electronic ions~\cite{Czarnecki2017,Czarnecki2016} may also contribute, and their
calculation can be extended to the case of muons in a straightforward manner.

In the calculation of the \textit{SE wave function correction} [Fig.~\ref{fig:QED1loop}~(c)], the muon propagator between the magnetic
interaction and the SE loop can be expressed as a spectral sum over all eigenfunctions $|n\rangle$ of the Coulomb-Dirac Hamiltonian as
\begin{equation}
\sum_{n} \frac{\vert n \rangle \langle n \vert}{E_a - E_n + {\rm sgn}(E_n) i0} \,,
\end{equation}
with the $E_n$ being the eigenenergies of the $\vert n \rangle$ and $E_a$ being the eigenenergy of the reference state $\vert a \rangle$.
The diagram needs to be split into the irreducible ($E_n \neq E_a$) and the reducible ($E_n = E_a$) part. The $g$-factor correction
of the irreducible part can be expressed using the SE operator $\Sigma$ as
\begin{equation}
\Delta g_{\rm SE,wf,irred} = - \frac{8 m_\mu}{3} \langle \delta a \vert \gamma^0 \Sigma (E_a) \vert a \rangle \,.
\end{equation}
Here, $\gamma^0$ is the time-like Dirac matrix. The irreducible part can be separated into the zero-potential contribution
(free internal muon line), the one-potential contribution (free internal muon line with one interaction with the nuclear potential)
and the many-potential contribution (two and more interactions with the nuclear potential). While the zero-potential and one-potential
contributions are UV divergent, the many-potential contribution is finite. The zero-potential contribution can be written as
\begin{equation}
\Delta g_{\rm SE,wf,irred}^{[0]} = - \frac{8 m_\mu}{3} \langle \delta a \vert \gamma^0 \Sigma_2 \vert a \rangle \,.
\end{equation}
Here, $\Sigma_2 (p)$ is the momentum-space SE function of the free muon and using dimensional regularization in $d=4-\epsilon$ dimensions,
it can be expressed as~\cite{Yerokhin1999}
\begin{equation}
\Sigma_2 (p) = \delta m - \frac{\alpha}{4 \pi} \Delta_\epsilon ( \slashed{p}  - m_{\mu} ) + \Sigma_{\rm R} (p) \,,
\end{equation}
with $\delta m = \frac{3 \alpha m_{\mu}}{4 \pi} \left( \Delta_\epsilon + \frac{4}{3} \right)$ and
$\Delta_\epsilon = \frac{2}{\epsilon} - \gamma_{\rm E} - \ln m_{\mu}^2 + \ln 4 \pi$, where $\gamma_{\rm E}=0.57721\dots$ is Euler's constant.
The $\delta m$ term is cancelled by mass renormalization, and the $\Delta_\epsilon$ term will be cancelled by a similar term in the
one-potential contribution. The renormalized zero-potential contribution is defined as
\begin{equation}
\Delta g_{\rm SE,wf,irred,ren}^{[0]} = - \frac{8 m_\mu}{3} \langle \delta a \vert \gamma^0 \Sigma_{\rm R} \vert a \rangle\,,
\end{equation}
while the one-potential contribution is
\begin{equation}
\Delta g_{\rm SE,wf,irred}^{[1]} = - \frac{8 m_\mu}{3} \langle \delta a \vert \gamma^0 \Gamma^0_2 V \vert a \rangle \,,
\end{equation}
with $V$ being the interaction potential of the nucleus. $\Gamma^{\nu}_2 (p',p)$ ($\nu \in \left\{0,1,2,3\right\}$) is the vertex function
for free fermions and can be separated into a divergent and a regular part as~\cite{Yerokhin1999}
\begin{equation}
\Gamma^{\nu}_2 (p',p) = \frac{\alpha}{4 \pi} \Delta_\epsilon \gamma^{\nu} + \Gamma^{\nu}_{\rm R} (p',p) \,.
\end{equation}
The $\Delta_\epsilon$ term in the one-potential contribution cancels the corresponding $\Delta_\epsilon$ term in the
zero-potential contribution. For details of the renormalization procedure and for expressions of $\Sigma_{\rm R}(p)$ and
$\Gamma^{\nu}_{\rm R}(p',p)$ see Ref.~\cite{Yerokhin1999}. The renormalized one-potential term is then defined as
\begin{equation}
\Delta g_{\rm SE,wf,irred,ren}^{[1]} = - \frac{8 m_\mu}{3} \langle \delta a \vert \gamma^0 \Gamma^0_{\rm R} V \vert a \rangle \,.
\end{equation}
The many-potential contribution was evaluated using methods described in Ref.~\cite{Johnson1988,Yerokhin1999}. It is
straightforward to generalize the calculation of the Lamb-shift diagram to the many-potential contribution of the $g$-factor
SE diagram. The integration over the virtual photon frequency required in the many-potential contribution was split into a
low-energy and a high-energy part. The partial-wave expansion of the low-energy part converges rapidly and does not require any
extrapolation. The high-energy term converges slower. The series was computed up to Dirac angular momentum quantum numbers
$\vert \kappa \vert \approx 40$, and the remainder of the series was estimated using the Richardson extrapolation
method~\cite{Richardson1911}.

\begin{table*}[t!]
\begin{ruledtabular}
\begin{tabular}{llll}
Effect                 & Term                 & Numerical value                & Ref. \\
\hline
Dirac value            &                      & \phantom{-}1.999 857 988 8     & \cite{Breit1928,CODATA2014} \\
Finite nuclear size    &                      & \phantom{-}0.000 000 094 6(4)  & \cite{Angeli2013} \\
One-loop SE            & $(Z \alpha)^0$       & \phantom{-}0.002 322 819 5     & \cite{Schwinger48,CODATA2014} \\
                       & all-order binding & \phantom{-}0.000 000 084 9(10) & \\
One-loop VP            & $e$VP, Uehling    & -0.000 000 479 6                 & \\
                       & $e$VP, magnetic loop & \phantom{-}0.000 000 127 2(4) & \\
                       & $\mu$VP, Uehling     &           -0.000 000 000 1    & \\
                       & hadronic VP, Uehling &           -0.000 000 000 1(1) & \\
Two-loop QED           &  $(Z \alpha)^0$      & \phantom{-}0.000 008 264 4    &  \cite{Peterman57,Sommerfield58} \\
                       & SE-SE, $(Z \alpha)^2$--- $(Z \alpha)^5$ & -0.000 000 000 1& \cite{Eides1997,Czarnecki2000,Pachucki2005,Czarnecki2017}\\
                       & S(eVP)E, $(Z \alpha)^2$                 & \phantom{-}0.000 000 000 4& \cite{Peterman57,Sommerfield58,Eides1997,Czarnecki2000}\\
                       & 2nd-order Uehling  & -0.000 000 001 1(4)               & \\
                       & K\"all\'en-Sabry      & -0.000 000 003 5               & \\
                       & magnetic loop+Uehling & \phantom{-}0.000 000 000 3     & \\
$\ge$ Three-loop QED   & $(Z \alpha)^0$        & \phantom{-}0.000 000 610 6     & \cite{Laporta96,Aoyama07,Aoyama12,CODATA2014} \\
Nuclear recoil         & $\left(\frac{m}{M}\right)^1$, all orders in $Z \alpha$ & \phantom{-}0.000 006 038 2 &  \cite{Shabaev2002}\\
                       & $\left(\frac{m}{M}\right)^{2+}$, $(Z \alpha)^2$ & -0.000 000 488 7 &  \cite{Pachucki2008}\\
                       & radiative recoil &  -0.000 000 004 7 & \cite{Grotch1970}\\
Weak interaction       & $(Z\alpha)^0$         &  \phantom{-}0.000 000 003 1     & \cite{Czarnecki96,CODATA2014} \\
Hadronic contributions &  $(Z\alpha)^0$        &  \phantom{-}0.000 000 139 3(12) & \cite{Prades10,Nomura13,Kurz14,CODATA2014} \\
\hline
Sum                    &                       &  \phantom{-}2.002 195 193 4(20) & \\
\end{tabular}
\caption{\label{gHe} Various contributions to the $g$ factor of $\mu{}^4$He$^+$. The abbreviations are:
\enquote{eVP}/\enquote{$\mu$VP}: VP due to virtual $e^-e^+$/$\mu^- \mu^+$ pairs.
The estimated uncertainty of the nuclear size effect stems from the error bar of the root-mean-square nuclear radius and the
uncertainty of the nuclear charge distribution model.
}
\end{ruledtabular}
\end{table*}

The $g$-factor contribution of the reducible SE diagram is calculated from the energy derivative of the Lamb shift matrix element:
\begin{equation}
\Delta g_{\rm red} = g_{\rm D} \left. \frac{\partial}{\partial E} \langle a \vert \gamma^0 \Sigma (E) \vert a \rangle \right|_{E=E_a} \,.
\end{equation}
It can be again split into the zero- and many-potential contributions. While the zero-potential contribution is UV divergent,
the one-potential part is finite and can therefore be included in the many-potential term.

The \textit{SE vertex correction} [Fig.~\ref{fig:QED1loop}~(d)] can be expressed as
$\Delta E_{\rm ver} = - e \langle a \vert \gamma^0 \boldsymbol{\Gamma} \cdot \mathbf{A} \vert a \rangle$, with $\boldsymbol{\Gamma}$
being the 3-vector component of the vertex function. This expression can be split into the zero- and many-potential contributions.
The zero-potential term is UV divergent, but the many-potential part does not contain UV divergences. The UV-divergent terms in the
zero-potential contributions of the vertex diagram and the reducible SE diagram cancel each other. The renormalized zero-potential
contribution can be calculated in momentum space, using the magnetic vector potential
$
\mathbf{A} ( \mathbf{p}' - \mathbf{p} ) = - \frac{i}{2} (2 \pi)^3 \mathbf{B} \times
\nabla_{\mathbf{p}} \delta ( \mathbf{p}' - \mathbf{p} )
$,
and can be expressed as~\cite{Yerokhin2004}
\begin{align}
\Delta g_{\rm ver}^{[0]} = & - 2 i m \int \frac{\mathsf{d}^3 p}{(2 \pi)^3} \int \mathsf{d}^3 p' \overline{a} (\mathbf{p}) \\
& \left( \nabla_{\mathbf{p}'} \delta ( \mathbf{p} - \mathbf{p}' ) \times \boldsymbol{\Gamma}_R (p,p') \right)_z a (\mathbf{p}')\,, \nonumber
\end{align}
with $a (\mathbf{p})$ being the muon wave function in momentum representation.
This can be further evaluated using integration by parts. For further details see Ref.~\cite{Yerokhin2004}. We note that our numerical results
for the SE terms agree well with $Z\alpha$ expansion formulas~\cite{Eides1997,Pachucki2004,Pachucki2005}.

The calculations so far have been performed in the Furry picture \cite{Furry1951}, i.e. using a static external field to describe
the nucleus. The \textit{nuclear recoil} contribution is the correction to the $g$ factor due to the finiteness of the nuclear mass.
Formulas derived for bound electrons \cite{Grotch1970,Shabaev2002,Pachucki2008} are applicable also to the case of muonic ions.

\textit{Results and conclusions --}
The highest theoretical accuracy can be achieved in light muonic ions, as all binding corrections scale with high powers of the
atomic number $Z$. Especially nuclear structural effects are suppressed, which, given the uncertainties of nuclear parameters,
is necessary for a sufficiently accurate competitive determination of $a_{\mu}$ or $m_{\mu}$. Therefore, we chose the muonic ion with
the lightest spinless nucleus, namely, $^4$He$^+$ for an illustration of our theory. Table~\ref{gHe} lists numerical results for the
contributing terms.

The binding effect on the $g$ factor for this element can be calculated with a $10^{-9}$ fractional accuracy, allowing for the
improvement of the muon mass, or a determination of the free muon magnetic moment anomaly by subtracting theoretical binding effects
from the measured bound-muon $g$ factor. Such experiments are challenging due to the short lifetime of the muon. Nevertheless, in light
of recent advances in the creation and precision spectroscopy of light muonic atoms~\cite{Pohl2010,Antognini2013,Pohl2017} and
Penning-trap techniques such as phase-sensitive cyclotron frequency measurements~\cite{Sturm2011b,Sturm2017}, this method may serve
in near future as an independent muon mass or magnetic moment anomaly determination technique, and along with corresponding experimental
developments, will improve the mass uncertainty by more than an order of magnitude.

\begin{acknowledgements}

We acknowledge insightful conversations with Jacek Zatorski. This work is part of and supported by the
German Research Foundation (DFG) Collaborative Research Centre "SFB 1225 (ISOQUANT)".
V.A.Y. acknowledges support by the Ministry of Education and Science of the Russian Federation
Grant No.~3.5397.2017/6.7.

\end{acknowledgements}


\begin{thebibliography}{64}
\expandafter\ifx\csname natexlab\endcsname\relax\def\natexlab#1{#1}\fi
\expandafter\ifx\csname bibnamefont\endcsname\relax
  \def\bibnamefont#1{#1}\fi
\expandafter\ifx\csname bibfnamefont\endcsname\relax
  \def\bibfnamefont#1{#1}\fi
\expandafter\ifx\csname citenamefont\endcsname\relax
  \def\citenamefont#1{#1}\fi
\expandafter\ifx\csname url\endcsname\relax
  \def\url#1{\texttt{#1}}\fi
\expandafter\ifx\csname urlprefix\endcsname\relax\def\urlprefix{URL }\fi
\providecommand{\bibinfo}[2]{#2}
\providecommand{\eprint}[2][]{\url{#2}}

\bibitem[{\citenamefont{Jegerlehner and Nyffeler}(2009)}]{Jegerlehner2009}
\bibinfo{author}{\bibfnamefont{F.}~\bibnamefont{Jegerlehner}} \bibnamefont{and}
  \bibinfo{author}{\bibfnamefont{A.}~\bibnamefont{Nyffeler}},
  \bibinfo{journal}{Phys. Rep.} \textbf{\bibinfo{volume}{477}},
  \bibinfo{pages}{1 } (\bibinfo{year}{2009}).

\bibitem[{\citenamefont{Pohl et~al.}(2010)\citenamefont{Pohl, Antognini, Nez,
  Amaro, Biraben, Cardoso, Covita, Dax, Dhawan, Fernandes et~al.}}]{Pohl2010}
\bibinfo{author}{\bibfnamefont{R.}~\bibnamefont{Pohl}},
  \bibinfo{author}{\bibfnamefont{A.}~\bibnamefont{Antognini}},
  \bibinfo{author}{\bibfnamefont{F.}~\bibnamefont{Nez}},
  \bibinfo{author}{\bibfnamefont{F.~D.} \bibnamefont{Amaro}},
  \bibinfo{author}{\bibfnamefont{F.}~\bibnamefont{Biraben}},
  \bibinfo{author}{\bibfnamefont{J.~a. M.~R.} \bibnamefont{Cardoso}},
  \bibinfo{author}{\bibfnamefont{D.~S.} \bibnamefont{Covita}},
  \bibinfo{author}{\bibfnamefont{A.}~\bibnamefont{Dax}},
  \bibinfo{author}{\bibfnamefont{S.}~\bibnamefont{Dhawan}},
  \bibinfo{author}{\bibfnamefont{L.~M.~P.} \bibnamefont{Fernandes}},
  \bibnamefont{et~al.}, \bibinfo{journal}{Nature}
  \textbf{\bibinfo{volume}{466}}, \bibinfo{pages}{213 } (\bibinfo{year}{2010}).

\bibitem[{\citenamefont{Antognini et~al.}(2013)\citenamefont{Antognini, Nez,
  Schuhmann, Amar, Biraben, Cardoso, Covita, Dax, Dhawan, Diepold
  et~al.}}]{Antognini2013}
\bibinfo{author}{\bibfnamefont{A.}~\bibnamefont{Antognini}},
  \bibinfo{author}{\bibfnamefont{F.}~\bibnamefont{Nez}},
  \bibinfo{author}{\bibfnamefont{K.}~\bibnamefont{Schuhmann}},
  \bibinfo{author}{\bibfnamefont{F.~D.} \bibnamefont{Amar}},
  \bibinfo{author}{\bibfnamefont{F.}~\bibnamefont{Biraben}},
  \bibinfo{author}{\bibfnamefont{J.~M.~R.} \bibnamefont{Cardoso}},
  \bibinfo{author}{\bibfnamefont{D.~S.} \bibnamefont{Covita}},
  \bibinfo{author}{\bibfnamefont{A.}~\bibnamefont{Dax}},
  \bibinfo{author}{\bibfnamefont{S.}~\bibnamefont{Dhawan}},
  \bibinfo{author}{\bibfnamefont{M.}~\bibnamefont{Diepold}},
  \bibnamefont{et~al.}, \bibinfo{journal}{Science}
  \textbf{\bibinfo{volume}{339}}, \bibinfo{pages}{417 } (\bibinfo{year}{2013}).

\bibitem[{\citenamefont{Beyer et~al.}(2017)\citenamefont{Beyer, Maisenbacher,
  Matveev, Pohl, Khabarova, Grinin, Lamour, Yost, H{\"a}nsch, Kolachevsky
  et~al.}}]{Beyer2017}
\bibinfo{author}{\bibfnamefont{A.}~\bibnamefont{Beyer}},
  \bibinfo{author}{\bibfnamefont{L.}~\bibnamefont{Maisenbacher}},
  \bibinfo{author}{\bibfnamefont{A.}~\bibnamefont{Matveev}},
  \bibinfo{author}{\bibfnamefont{R.}~\bibnamefont{Pohl}},
  \bibinfo{author}{\bibfnamefont{K.}~\bibnamefont{Khabarova}},
  \bibinfo{author}{\bibfnamefont{A.}~\bibnamefont{Grinin}},
  \bibinfo{author}{\bibfnamefont{T.}~\bibnamefont{Lamour}},
  \bibinfo{author}{\bibfnamefont{D.~C.} \bibnamefont{Yost}},
  \bibinfo{author}{\bibfnamefont{T.~W.} \bibnamefont{H{\"a}nsch}},
  \bibinfo{author}{\bibfnamefont{N.}~\bibnamefont{Kolachevsky}},
  \bibnamefont{et~al.}, \bibinfo{journal}{Science}
  \textbf{\bibinfo{volume}{358}}, \bibinfo{pages}{79} (\bibinfo{year}{2017}).

\bibitem[{\citenamefont{H\"affner et~al.}(2000)\citenamefont{H\"affner, Beier,
  Hermanspahn, Kluge, Quint, Stahl, Verd\'u, and Werth}}]{Haffner2000}
\bibinfo{author}{\bibfnamefont{H.}~\bibnamefont{H\"affner}},
  \bibinfo{author}{\bibfnamefont{T.}~\bibnamefont{Beier}},
  \bibinfo{author}{\bibfnamefont{N.}~\bibnamefont{Hermanspahn}},
  \bibinfo{author}{\bibfnamefont{H.-J.} \bibnamefont{Kluge}},
  \bibinfo{author}{\bibfnamefont{W.}~\bibnamefont{Quint}},
  \bibinfo{author}{\bibfnamefont{S.}~\bibnamefont{Stahl}},
  \bibinfo{author}{\bibfnamefont{J.}~\bibnamefont{Verd\'u}}, \bibnamefont{and}
  \bibinfo{author}{\bibfnamefont{G.}~\bibnamefont{Werth}},
  \bibinfo{journal}{Phys. Rev. Lett.} \textbf{\bibinfo{volume}{85}},
  \bibinfo{pages}{5308} (\bibinfo{year}{2000}).

\bibitem[{\citenamefont{Verd\'u et~al.}(2004)\citenamefont{Verd\'u,
  Djeki\ifmmode~\acute{c}\else \'{c}\fi{}, Stahl, Valenzuela, Vogel, Werth,
  Beier, Kluge, and Quint}}]{Verdu2004}
\bibinfo{author}{\bibfnamefont{J.}~\bibnamefont{Verd\'u}},
  \bibinfo{author}{\bibfnamefont{S.}~\bibnamefont{Djeki\ifmmode~\acute{c}\else
  \'{c}\fi{}}}, \bibinfo{author}{\bibfnamefont{S.}~\bibnamefont{Stahl}},
  \bibinfo{author}{\bibfnamefont{T.}~\bibnamefont{Valenzuela}},
  \bibinfo{author}{\bibfnamefont{M.}~\bibnamefont{Vogel}},
  \bibinfo{author}{\bibfnamefont{G.}~\bibnamefont{Werth}},
  \bibinfo{author}{\bibfnamefont{T.}~\bibnamefont{Beier}},
  \bibinfo{author}{\bibfnamefont{H.-J.} \bibnamefont{Kluge}}, \bibnamefont{and}
  \bibinfo{author}{\bibfnamefont{W.}~\bibnamefont{Quint}},
  \bibinfo{journal}{Phys. Rev. Lett.} \textbf{\bibinfo{volume}{92}},
  \bibinfo{pages}{093002} (\bibinfo{year}{2004}).

\bibitem[{\citenamefont{Sturm et~al.}(2011{\natexlab{a}})\citenamefont{Sturm,
  Wagner, Schabinger, Zatorski, Harman, Quint, Werth, Keitel, and
  Blaum}}]{Sturm2011}
\bibinfo{author}{\bibfnamefont{S.}~\bibnamefont{Sturm}},
  \bibinfo{author}{\bibfnamefont{A.}~\bibnamefont{Wagner}},
  \bibinfo{author}{\bibfnamefont{B.}~\bibnamefont{Schabinger}},
  \bibinfo{author}{\bibfnamefont{J.}~\bibnamefont{Zatorski}},
  \bibinfo{author}{\bibfnamefont{Z.}~\bibnamefont{Harman}},
  \bibinfo{author}{\bibfnamefont{W.}~\bibnamefont{Quint}},
  \bibinfo{author}{\bibfnamefont{G.}~\bibnamefont{Werth}},
  \bibinfo{author}{\bibfnamefont{C.~H.} \bibnamefont{Keitel}},
  \bibnamefont{and} \bibinfo{author}{\bibfnamefont{K.}~\bibnamefont{Blaum}},
  \bibinfo{journal}{Phys. Rev. Lett.} \textbf{\bibinfo{volume}{107}},
  \bibinfo{pages}{023002} (\bibinfo{year}{2011}{\natexlab{a}}).

\bibitem[{\citenamefont{Sturm et~al.}(2011{\natexlab{b}})\citenamefont{Sturm,
  Wagner, Schabinger, and Blaum}}]{Sturm2011b}
\bibinfo{author}{\bibfnamefont{S.}~\bibnamefont{Sturm}},
  \bibinfo{author}{\bibfnamefont{A.}~\bibnamefont{Wagner}},
  \bibinfo{author}{\bibfnamefont{B.}~\bibnamefont{Schabinger}},
  \bibnamefont{and} \bibinfo{author}{\bibfnamefont{K.}~\bibnamefont{Blaum}},
  \bibinfo{journal}{Phys. Rev. Lett.} \textbf{\bibinfo{volume}{107}},
  \bibinfo{pages}{143003} (\bibinfo{year}{2011}{\natexlab{b}}).

\bibitem[{\citenamefont{Beier}(2000)}]{BeierReport}
\bibinfo{author}{\bibfnamefont{T.}~\bibnamefont{Beier}},
  \bibinfo{journal}{Phys. Rep.} \textbf{\bibinfo{volume}{339}},
  \bibinfo{pages}{79} (\bibinfo{year}{2000}).

\bibitem[{\citenamefont{Yerokhin et~al.}(2004)\citenamefont{Yerokhin,
  Indelicato, and Shabaev}}]{Yerokhin2004}
\bibinfo{author}{\bibfnamefont{V.~A.} \bibnamefont{Yerokhin}},
  \bibinfo{author}{\bibfnamefont{P.}~\bibnamefont{Indelicato}},
  \bibnamefont{and} \bibinfo{author}{\bibfnamefont{V.~M.}
  \bibnamefont{Shabaev}}, \bibinfo{journal}{Phys. Rev. A}
  \textbf{\bibinfo{volume}{69}}, \bibinfo{pages}{052503}
  (\bibinfo{year}{2004}).

\bibitem[{\citenamefont{Pachucki et~al.}(2004)\citenamefont{Pachucki,
  Jentschura, and Yerokhin}}]{Pachucki2004}
\bibinfo{author}{\bibfnamefont{K.}~\bibnamefont{Pachucki}},
  \bibinfo{author}{\bibfnamefont{U.~D.} \bibnamefont{Jentschura}},
  \bibnamefont{and} \bibinfo{author}{\bibfnamefont{V.~A.}
  \bibnamefont{Yerokhin}}, \bibinfo{journal}{Phys. Rev. Lett.}
  \textbf{\bibinfo{volume}{93}}, \bibinfo{pages}{150401}
  (\bibinfo{year}{2004}).

\bibitem[{\citenamefont{Pachucki and Puchalski}(2017)}]{Pachucki2017}
\bibinfo{author}{\bibfnamefont{K.}~\bibnamefont{Pachucki}} \bibnamefont{and}
  \bibinfo{author}{\bibfnamefont{M.}~\bibnamefont{Puchalski}},
  \bibinfo{journal}{Phys. Rev. A} \textbf{\bibinfo{volume}{96}},
  \bibinfo{pages}{032503} (\bibinfo{year}{2017}).

\bibitem[{\citenamefont{Czarnecki et~al.}(2017)\citenamefont{Czarnecki,
  Dowling, Piclum, and Szafron}}]{Czarnecki2017}
\bibinfo{author}{\bibfnamefont{A.}~\bibnamefont{Czarnecki}},
  \bibinfo{author}{\bibfnamefont{M.}~\bibnamefont{Dowling}},
  \bibinfo{author}{\bibfnamefont{J.}~\bibnamefont{Piclum}}, \bibnamefont{and}
  \bibinfo{author}{\bibfnamefont{R.}~\bibnamefont{Szafron}},
  \bibinfo{journal}{arXiv preprint arXiv:1711.00190}  (\bibinfo{year}{2017}).

\bibitem[{\citenamefont{Beier et~al.}(2001)\citenamefont{Beier, H\"affner,
  Hermanspahn, Karshenboim, Kluge, Quint, Stahl, Verd\'u, and
  Werth}}]{Beier2001}
\bibinfo{author}{\bibfnamefont{T.}~\bibnamefont{Beier}},
  \bibinfo{author}{\bibfnamefont{H.}~\bibnamefont{H\"affner}},
  \bibinfo{author}{\bibfnamefont{N.}~\bibnamefont{Hermanspahn}},
  \bibinfo{author}{\bibfnamefont{S.~G.} \bibnamefont{Karshenboim}},
  \bibinfo{author}{\bibfnamefont{H.-J.} \bibnamefont{Kluge}},
  \bibinfo{author}{\bibfnamefont{W.}~\bibnamefont{Quint}},
  \bibinfo{author}{\bibfnamefont{S.}~\bibnamefont{Stahl}},
  \bibinfo{author}{\bibfnamefont{J.}~\bibnamefont{Verd\'u}}, \bibnamefont{and}
  \bibinfo{author}{\bibfnamefont{G.}~\bibnamefont{Werth}},
  \bibinfo{journal}{Phys. Rev. Lett.} \textbf{\bibinfo{volume}{88}},
  \bibinfo{pages}{011603} (\bibinfo{year}{2001}).

\bibitem[{\citenamefont{Sturm et~al.}(2014)\citenamefont{Sturm, K{\"o}hler,
  Zatorski, Wagner, Harman, Werth, Quint, Keitel, and Blaum}}]{Sturm2014}
\bibinfo{author}{\bibfnamefont{S.}~\bibnamefont{Sturm}},
  \bibinfo{author}{\bibfnamefont{F.}~\bibnamefont{K{\"o}hler}},
  \bibinfo{author}{\bibfnamefont{J.}~\bibnamefont{Zatorski}},
  \bibinfo{author}{\bibfnamefont{A.}~\bibnamefont{Wagner}},
  \bibinfo{author}{\bibfnamefont{Z.}~\bibnamefont{Harman}},
  \bibinfo{author}{\bibfnamefont{G.}~\bibnamefont{Werth}},
  \bibinfo{author}{\bibfnamefont{W.}~\bibnamefont{Quint}},
  \bibinfo{author}{\bibfnamefont{C.~H.} \bibnamefont{Keitel}},
  \bibnamefont{and} \bibinfo{author}{\bibfnamefont{K.}~\bibnamefont{Blaum}},
  \bibinfo{journal}{Nature} \textbf{\bibinfo{volume}{506}},
  \bibinfo{pages}{467} (\bibinfo{year}{2014}).

\bibitem[{\citenamefont{K\"ohler et~al.}(2015)\citenamefont{K\"ohler, Sturm,
  Kracke, Werth, Quint, and Blaum}}]{Kohler15}
\bibinfo{author}{\bibfnamefont{F.}~\bibnamefont{K\"ohler}},
  \bibinfo{author}{\bibfnamefont{S.}~\bibnamefont{Sturm}},
  \bibinfo{author}{\bibfnamefont{A.}~\bibnamefont{Kracke}},
  \bibinfo{author}{\bibfnamefont{G.}~\bibnamefont{Werth}},
  \bibinfo{author}{\bibfnamefont{W.}~\bibnamefont{Quint}}, \bibnamefont{and}
  \bibinfo{author}{\bibfnamefont{K.}~\bibnamefont{Blaum}}, \bibinfo{journal}{J.
  Phys. B} \textbf{\bibinfo{volume}{48}}, \bibinfo{pages}{144032}
  (\bibinfo{year}{2015}).

\bibitem[{\citenamefont{Zatorski et~al.}(2017)\citenamefont{Zatorski, Sikora,
  Karshenboim, Sturm, K\"ohler-Langes, Blaum, Keitel, and
  Harman}}]{Zatorski2017}
\bibinfo{author}{\bibfnamefont{J.}~\bibnamefont{Zatorski}},
  \bibinfo{author}{\bibfnamefont{B.}~\bibnamefont{Sikora}},
  \bibinfo{author}{\bibfnamefont{S.~G.} \bibnamefont{Karshenboim}},
  \bibinfo{author}{\bibfnamefont{S.}~\bibnamefont{Sturm}},
  \bibinfo{author}{\bibfnamefont{F.}~\bibnamefont{K\"ohler-Langes}},
  \bibinfo{author}{\bibfnamefont{K.}~\bibnamefont{Blaum}},
  \bibinfo{author}{\bibfnamefont{C.~H.} \bibnamefont{Keitel}},
  \bibnamefont{and} \bibinfo{author}{\bibfnamefont{Z.}~\bibnamefont{Harman}},
  \bibinfo{journal}{Phys. Rev. A} \textbf{\bibinfo{volume}{96}},
  \bibinfo{pages}{012502} (\bibinfo{year}{2017}).

\bibitem[{\citenamefont{Jungmann}(2000)}]{Jungmann2000}
\bibinfo{author}{\bibfnamefont{K.}~\bibnamefont{Jungmann}},
  \bibinfo{journal}{Hyperfine Interact.} \textbf{\bibinfo{volume}{127}},
  \bibinfo{pages}{189} (\bibinfo{year}{2000}), ISSN \bibinfo{issn}{0304-3843}.

\bibitem[{\citenamefont{Mohr et~al.}(2016{\natexlab{a}})\citenamefont{Mohr,
  Newell, and Taylor}}]{CODATA2014}
\bibinfo{author}{\bibfnamefont{P.~J.} \bibnamefont{Mohr}},
  \bibinfo{author}{\bibfnamefont{D.~B.} \bibnamefont{Newell}},
  \bibnamefont{and} \bibinfo{author}{\bibfnamefont{B.~N.}
  \bibnamefont{Taylor}}, \bibinfo{journal}{Rev. Mod. Phys.}
  \textbf{\bibinfo{volume}{88}}, \bibinfo{pages}{035009}
  (\bibinfo{year}{2016}{\natexlab{a}}).

\bibitem[{\citenamefont{Mohr et~al.}(2016{\natexlab{b}})\citenamefont{Mohr,
  Newell, and Taylor}}]{CODATA2014b}
\bibinfo{author}{\bibfnamefont{P.~J.} \bibnamefont{Mohr}},
  \bibinfo{author}{\bibfnamefont{D.~B.} \bibnamefont{Newell}},
  \bibnamefont{and} \bibinfo{author}{\bibfnamefont{B.~N.}
  \bibnamefont{Taylor}}, \bibinfo{journal}{J. Phys. Chem. Ref. Data}
  \textbf{\bibinfo{volume}{45}}, \bibinfo{pages}{043102}
  (\bibinfo{year}{2016}{\natexlab{b}}).

\bibitem[{\citenamefont{Breit}(1928)}]{Breit1928}
\bibinfo{author}{\bibfnamefont{G.}~\bibnamefont{Breit}},
  \bibinfo{journal}{Nature} \textbf{\bibinfo{volume}{122}},
  \bibinfo{pages}{649} (\bibinfo{year}{1928}).

\bibitem[{\citenamefont{Karshenboim}(2000)}]{Karshenboim2000}
\bibinfo{author}{\bibfnamefont{S.~G.} \bibnamefont{Karshenboim}},
  \bibinfo{journal}{Phys. Lett. A} \textbf{\bibinfo{volume}{266}},
  \bibinfo{pages}{380 } (\bibinfo{year}{2000}).

\bibitem[{\citenamefont{Zatorski et~al.}(2012)\citenamefont{Zatorski,
  Oreshkina, Keitel, and Harman}}]{Zatorski2012}
\bibinfo{author}{\bibfnamefont{J.}~\bibnamefont{Zatorski}},
  \bibinfo{author}{\bibfnamefont{N.~S.} \bibnamefont{Oreshkina}},
  \bibinfo{author}{\bibfnamefont{C.~H.} \bibnamefont{Keitel}},
  \bibnamefont{and} \bibinfo{author}{\bibfnamefont{Z.}~\bibnamefont{Harman}},
  \bibinfo{journal}{Phys. Rev. Lett.} \textbf{\bibinfo{volume}{108}},
  \bibinfo{pages}{063005} (\bibinfo{year}{2012}).

\bibitem[{\citenamefont{Frosch et~al.}(1967)\citenamefont{Frosch, McCarthy,
  Rand, and Yearian}}]{Frosch1967}
\bibinfo{author}{\bibfnamefont{R.~F.} \bibnamefont{Frosch}},
  \bibinfo{author}{\bibfnamefont{J.~S.} \bibnamefont{McCarthy}},
  \bibinfo{author}{\bibfnamefont{R.~E.} \bibnamefont{Rand}}, \bibnamefont{and}
  \bibinfo{author}{\bibfnamefont{M.~R.} \bibnamefont{Yearian}},
  \bibinfo{journal}{Phys. Rev.} \textbf{\bibinfo{volume}{160}},
  \bibinfo{pages}{874} (\bibinfo{year}{1967}).

\bibitem[{\citenamefont{Nefiodov et~al.}(2002)\citenamefont{Nefiodov, Plunien,
  and Soff}}]{Nefiodov2002}
\bibinfo{author}{\bibfnamefont{A.~V.} \bibnamefont{Nefiodov}},
  \bibinfo{author}{\bibfnamefont{G.}~\bibnamefont{Plunien}}, \bibnamefont{and}
  \bibinfo{author}{\bibfnamefont{G.}~\bibnamefont{Soff}},
  \bibinfo{journal}{Phys. Rev. Lett.} \textbf{\bibinfo{volume}{89}},
  \bibinfo{pages}{081802} (\bibinfo{year}{2002}).

\bibitem[{\citenamefont{Volotka and Plunien}(2014)}]{Volotka2014}
\bibinfo{author}{\bibfnamefont{A.~V.} \bibnamefont{Volotka}} \bibnamefont{and}
  \bibinfo{author}{\bibfnamefont{G.}~\bibnamefont{Plunien}},
  \bibinfo{journal}{Phys. Rev. Lett.} \textbf{\bibinfo{volume}{113}},
  \bibinfo{pages}{023002} (\bibinfo{year}{2014}).

\bibitem[{\citenamefont{Shabaev}(2002)}]{Shabaev2002Report}
\bibinfo{author}{\bibfnamefont{V.~M.} \bibnamefont{Shabaev}},
  \bibinfo{journal}{Phys. Rep.} \textbf{\bibinfo{volume}{356}},
  \bibinfo{pages}{119 } (\bibinfo{year}{2002}).

\bibitem[{\citenamefont{Uehling}(1935)}]{Uehling1935}
\bibinfo{author}{\bibfnamefont{E.~A.} \bibnamefont{Uehling}},
  \bibinfo{journal}{Phys.~Rev.} \textbf{\bibinfo{volume}{48}},
  \bibinfo{pages}{55} (\bibinfo{year}{1935}).

\bibitem[{\citenamefont{Shabaev}(2003)}]{Shabaev2003}
\bibinfo{author}{\bibfnamefont{V.~M.} \bibnamefont{Shabaev}}, in
  \emph{\bibinfo{booktitle}{Precision Physics of Simple Atomic Systems}},
  edited by \bibinfo{editor}{\bibfnamefont{S.~G.} \bibnamefont{Karshenboim}}
  \bibnamefont{and} \bibinfo{editor}{\bibfnamefont{V.~B.}
  \bibnamefont{Smirnov}} (\bibinfo{publisher}{Springer},
  \bibinfo{address}{Berlin}, \bibinfo{year}{2003}), Lecture Notes in Physics,
  pp. \bibinfo{pages}{97 -- 113}.

\bibitem[{\citenamefont{Karshenboim et~al.}(2005)\citenamefont{Karshenboim,
  Lee, and Milstein}}]{Karshenboim2005}
\bibinfo{author}{\bibfnamefont{S.~G.} \bibnamefont{Karshenboim}},
  \bibinfo{author}{\bibfnamefont{R.~N.} \bibnamefont{Lee}}, \bibnamefont{and}
  \bibinfo{author}{\bibfnamefont{A.~I.} \bibnamefont{Milstein}},
  \bibinfo{journal}{Phys. Rev. A} \textbf{\bibinfo{volume}{72}},
  \bibinfo{pages}{042101} (\bibinfo{year}{2005}).

\bibitem[{\citenamefont{Karshenboim et~al.}(2001)\citenamefont{Karshenboim,
  Ivanov, and Shabaev}}]{Karshenboim2001}
\bibinfo{author}{\bibfnamefont{S.~G.} \bibnamefont{Karshenboim}},
  \bibinfo{author}{\bibfnamefont{V.~G.} \bibnamefont{Ivanov}},
  \bibnamefont{and} \bibinfo{author}{\bibfnamefont{V.~M.}
  \bibnamefont{Shabaev}}, \bibinfo{journal}{J. Exp. Theor. Phys. Lett.}
  \textbf{\bibinfo{volume}{93}}, \bibinfo{pages}{477} (\bibinfo{year}{2001}).

\bibitem[{\citenamefont{Fainshtein et~al.}(1991)\citenamefont{Fainshtein,
  Manakov, and Nekipelov}}]{Fainshtein1991}
\bibinfo{author}{\bibfnamefont{A.~G.} \bibnamefont{Fainshtein}},
  \bibinfo{author}{\bibfnamefont{N.~L.} \bibnamefont{Manakov}},
  \bibnamefont{and} \bibinfo{author}{\bibfnamefont{A.~A.}
  \bibnamefont{Nekipelov}}, \bibinfo{journal}{J. Phys. B}
  \textbf{\bibinfo{volume}{24}}, \bibinfo{pages}{559} (\bibinfo{year}{1991}).

\bibitem[{\citenamefont{Friar et~al.}(1999)\citenamefont{Friar, Martorell, and
  Sprung}}]{Friar1999}
\bibinfo{author}{\bibfnamefont{J.~L.} \bibnamefont{Friar}},
  \bibinfo{author}{\bibfnamefont{J.}~\bibnamefont{Martorell}},
  \bibnamefont{and} \bibinfo{author}{\bibfnamefont{D.~W.~L.}
  \bibnamefont{Sprung}}, \bibinfo{journal}{Phys. Rev. A}
  \textbf{\bibinfo{volume}{59}}, \bibinfo{pages}{4061} (\bibinfo{year}{1999}).

\bibitem[{\citenamefont{Shabaev et~al.}(2004)\citenamefont{Shabaev, Tupitsyn,
  Yerokhin, Plunien, and Soff}}]{Shabaev2004}
\bibinfo{author}{\bibfnamefont{V.~M.} \bibnamefont{Shabaev}},
  \bibinfo{author}{\bibfnamefont{I.~I.} \bibnamefont{Tupitsyn}},
  \bibinfo{author}{\bibfnamefont{V.~A.} \bibnamefont{Yerokhin}},
  \bibinfo{author}{\bibfnamefont{G.}~\bibnamefont{Plunien}}, \bibnamefont{and}
  \bibinfo{author}{\bibfnamefont{G.}~\bibnamefont{Soff}},
  \bibinfo{journal}{Phys. Rev. Lett.} \textbf{\bibinfo{volume}{93}},
  \bibinfo{pages}{130405} (\bibinfo{year}{2004}).

\bibitem[{\citenamefont{Michel et~al.}(2017)\citenamefont{Michel, Oreshkina,
  and Keitel}}]{Michel2017}
\bibinfo{author}{\bibfnamefont{N.}~\bibnamefont{Michel}},
  \bibinfo{author}{\bibfnamefont{N.~S.} \bibnamefont{Oreshkina}},
  \bibnamefont{and} \bibinfo{author}{\bibfnamefont{C.~H.}
  \bibnamefont{Keitel}}, \bibinfo{journal}{Phys. Rev. A}
  \textbf{\bibinfo{volume}{96}}, \bibinfo{pages}{032510}
  (\bibinfo{year}{2017}).

\bibitem[{\citenamefont{K{\"a}ll{\'e}n and Sabry}(1955)}]{Kaellen1955}
\bibinfo{author}{\bibfnamefont{G.}~\bibnamefont{K{\"a}ll{\'e}n}}
  \bibnamefont{and} \bibinfo{author}{\bibfnamefont{A.}~\bibnamefont{Sabry}},
  \bibinfo{journal}{K. Dan. Vidensk. Selsk. Mat. Fys. Medd.}
  \textbf{\bibinfo{volume}{29}} (\bibinfo{year}{1955}).

\bibitem[{\citenamefont{Fullerton and Rinker}(1976)}]{Fullerton1976}
\bibinfo{author}{\bibfnamefont{L.~W.} \bibnamefont{Fullerton}}
  \bibnamefont{and} \bibinfo{author}{\bibfnamefont{G.~A.}
  \bibnamefont{Rinker}}, \bibinfo{journal}{Phys. Rev. A}
  \textbf{\bibinfo{volume}{13}}, \bibinfo{pages}{1283} (\bibinfo{year}{1976}).

\bibitem[{\citenamefont{Indelicato}(2013)}]{Indelicato2013}
\bibinfo{author}{\bibfnamefont{P.}~\bibnamefont{Indelicato}},
  \bibinfo{journal}{Phys. Rev. A} \textbf{\bibinfo{volume}{87}},
  \bibinfo{pages}{022501} (\bibinfo{year}{2013}).

\bibitem[{\citenamefont{Lee et~al.}(2005)\citenamefont{Lee, Milstein, Terekhov,
  and Karshenboim}}]{Lee2005}
\bibinfo{author}{\bibfnamefont{R.~N.} \bibnamefont{Lee}},
  \bibinfo{author}{\bibfnamefont{A.~I.} \bibnamefont{Milstein}},
  \bibinfo{author}{\bibfnamefont{I.~S.} \bibnamefont{Terekhov}},
  \bibnamefont{and} \bibinfo{author}{\bibfnamefont{S.~G.}
  \bibnamefont{Karshenboim}}, \bibinfo{journal}{Phys. Rev. A}
  \textbf{\bibinfo{volume}{71}}, \bibinfo{pages}{052501}
  (\bibinfo{year}{2005}).

\bibitem[{\citenamefont{Lee et~al.}(2007)\citenamefont{Lee, Milstein, Terekhov,
  and Karshenboim}}]{Lee2007}
\bibinfo{author}{\bibfnamefont{R.~N.} \bibnamefont{Lee}},
  \bibinfo{author}{\bibfnamefont{A.~I.} \bibnamefont{Milstein}},
  \bibinfo{author}{\bibfnamefont{I.~S.} \bibnamefont{Terekhov}},
  \bibnamefont{and} \bibinfo{author}{\bibfnamefont{S.~G.}
  \bibnamefont{Karshenboim}}, \bibinfo{journal}{Can. J. Phys.}
  \textbf{\bibinfo{volume}{85}}, \bibinfo{pages}{541} (\bibinfo{year}{2007}).

\bibitem[{\citenamefont{Czarnecki and Szafron}(2016)}]{Czarnecki2016}
\bibinfo{author}{\bibfnamefont{A.}~\bibnamefont{Czarnecki}} \bibnamefont{and}
  \bibinfo{author}{\bibfnamefont{R.}~\bibnamefont{Szafron}},
  \bibinfo{journal}{Phys. Rev. A} \textbf{\bibinfo{volume}{94}},
  \bibinfo{pages}{060501} (\bibinfo{year}{2016}).

\bibitem[{\citenamefont{Yerokhin and Shabaev}(1999)}]{Yerokhin1999}
\bibinfo{author}{\bibfnamefont{V.~A.} \bibnamefont{Yerokhin}} \bibnamefont{and}
  \bibinfo{author}{\bibfnamefont{V.~M.} \bibnamefont{Shabaev}},
  \bibinfo{journal}{Phys. Rev. A} \textbf{\bibinfo{volume}{60}},
  \bibinfo{pages}{800} (\bibinfo{year}{1999}).

\bibitem[{\citenamefont{Johnson et~al.}(1988)\citenamefont{Johnson, Blundell,
  and Sapirstein}}]{Johnson1988}
\bibinfo{author}{\bibfnamefont{W.~R.} \bibnamefont{Johnson}},
  \bibinfo{author}{\bibfnamefont{S.~A.} \bibnamefont{Blundell}},
  \bibnamefont{and}
  \bibinfo{author}{\bibfnamefont{J.}~\bibnamefont{Sapirstein}},
  \bibinfo{journal}{Phys. Rev. A} \textbf{\bibinfo{volume}{37}},
  \bibinfo{pages}{2764} (\bibinfo{year}{1988}).

\bibitem[{\citenamefont{Richardson}(1911)}]{Richardson1911}
\bibinfo{author}{\bibfnamefont{L.~F.} \bibnamefont{Richardson}},
  \bibinfo{journal}{Philos. Trans. Royal Soc. A}
  \textbf{\bibinfo{volume}{210}}, \bibinfo{pages}{307} (\bibinfo{year}{1911}).

\bibitem[{\citenamefont{Angeli and Marinova}(2013)}]{Angeli2013}
\bibinfo{author}{\bibfnamefont{I.}~\bibnamefont{Angeli}} \bibnamefont{and}
  \bibinfo{author}{\bibfnamefont{K.}~\bibnamefont{Marinova}},
  \bibinfo{journal}{At. Data Nucl. Data Tables} \textbf{\bibinfo{volume}{99}},
  \bibinfo{pages}{69 } (\bibinfo{year}{2013}).

\bibitem[{\citenamefont{Schwinger}(1948)}]{Schwinger48}
\bibinfo{author}{\bibfnamefont{J.}~\bibnamefont{Schwinger}},
  \bibinfo{journal}{Phys. Rev.} \textbf{\bibinfo{volume}{73}},
  \bibinfo{pages}{416} (\bibinfo{year}{1948}).

\bibitem[{\citenamefont{Peterman}(1957)}]{Peterman57}
\bibinfo{author}{\bibfnamefont{A.}~\bibnamefont{Peterman}},
  \bibinfo{journal}{Helv. Phys. Act} \textbf{\bibinfo{volume}{30}},
  \bibinfo{pages}{407} (\bibinfo{year}{1957}).

\bibitem[{\citenamefont{Sommerfield}(1958)}]{Sommerfield58}
\bibinfo{author}{\bibfnamefont{C.~M.} \bibnamefont{Sommerfield}},
  \bibinfo{journal}{Ann. Phys.} \textbf{\bibinfo{volume}{5}},
  \bibinfo{pages}{26 } (\bibinfo{year}{1958}).

\bibitem[{\citenamefont{Eides and Grotch}(1997)}]{Eides1997}
\bibinfo{author}{\bibfnamefont{M.~I.} \bibnamefont{Eides}} \bibnamefont{and}
  \bibinfo{author}{\bibfnamefont{H.}~\bibnamefont{Grotch}},
  \bibinfo{journal}{Ann. Phys.} \textbf{\bibinfo{volume}{260}},
  \bibinfo{pages}{191 } (\bibinfo{year}{1997}).

\bibitem[{\citenamefont{Czarnecki et~al.}(2000)\citenamefont{Czarnecki,
  Melnikov, and Yelkhovsky}}]{Czarnecki2000}
\bibinfo{author}{\bibfnamefont{A.}~\bibnamefont{Czarnecki}},
  \bibinfo{author}{\bibfnamefont{K.}~\bibnamefont{Melnikov}}, \bibnamefont{and}
  \bibinfo{author}{\bibfnamefont{A.}~\bibnamefont{Yelkhovsky}},
  \bibinfo{journal}{Phys. Rev. A} \textbf{\bibinfo{volume}{63}},
  \bibinfo{pages}{012509} (\bibinfo{year}{2000}).

\bibitem[{\citenamefont{Pachucki et~al.}(2005)\citenamefont{Pachucki,
  Czarnecki, Jentschura, and Yerokhin}}]{Pachucki2005}
\bibinfo{author}{\bibfnamefont{K.}~\bibnamefont{Pachucki}},
  \bibinfo{author}{\bibfnamefont{A.}~\bibnamefont{Czarnecki}},
  \bibinfo{author}{\bibfnamefont{U.~D.} \bibnamefont{Jentschura}},
  \bibnamefont{and} \bibinfo{author}{\bibfnamefont{V.~A.}
  \bibnamefont{Yerokhin}}, \bibinfo{journal}{Phys. Rev. A}
  \textbf{\bibinfo{volume}{72}}, \bibinfo{pages}{022108}
  (\bibinfo{year}{2005}).

\bibitem[{\citenamefont{Laporta and Remiddi}(1996)}]{Laporta96}
\bibinfo{author}{\bibfnamefont{S.}~\bibnamefont{Laporta}} \bibnamefont{and}
  \bibinfo{author}{\bibfnamefont{E.}~\bibnamefont{Remiddi}},
  \bibinfo{journal}{Phys. Lett. B} \textbf{\bibinfo{volume}{379}},
  \bibinfo{pages}{283 } (\bibinfo{year}{1996}).

\bibitem[{\citenamefont{Aoyama et~al.}(2007)\citenamefont{Aoyama, Hayakawa,
  Kinoshita, and Nio}}]{Aoyama07}
\bibinfo{author}{\bibfnamefont{T.}~\bibnamefont{Aoyama}},
  \bibinfo{author}{\bibfnamefont{M.}~\bibnamefont{Hayakawa}},
  \bibinfo{author}{\bibfnamefont{T.}~\bibnamefont{Kinoshita}},
  \bibnamefont{and} \bibinfo{author}{\bibfnamefont{M.}~\bibnamefont{Nio}},
  \bibinfo{journal}{Phys. Rev. Lett.} \textbf{\bibinfo{volume}{99}},
  \bibinfo{pages}{110406} (\bibinfo{year}{2007}).

\bibitem[{\citenamefont{Aoyama et~al.}(2012)\citenamefont{Aoyama, Hayakawa,
  Kinoshita, and Nio}}]{Aoyama12}
\bibinfo{author}{\bibfnamefont{T.}~\bibnamefont{Aoyama}},
  \bibinfo{author}{\bibfnamefont{M.}~\bibnamefont{Hayakawa}},
  \bibinfo{author}{\bibfnamefont{T.}~\bibnamefont{Kinoshita}},
  \bibnamefont{and} \bibinfo{author}{\bibfnamefont{M.}~\bibnamefont{Nio}},
  \bibinfo{journal}{Phys. Rev. Lett.} \textbf{\bibinfo{volume}{109}},
  \bibinfo{pages}{111807} (\bibinfo{year}{2012}).

\bibitem[{\citenamefont{Shabaev and Yerokhin}(2002)}]{Shabaev2002}
\bibinfo{author}{\bibfnamefont{V.~M.} \bibnamefont{Shabaev}} \bibnamefont{and}
  \bibinfo{author}{\bibfnamefont{V.~A.} \bibnamefont{Yerokhin}},
  \bibinfo{journal}{Phys. Rev. Lett.} \textbf{\bibinfo{volume}{88}},
  \bibinfo{pages}{091801} (\bibinfo{year}{2002}).

\bibitem[{\citenamefont{Pachucki}(2008)}]{Pachucki2008}
\bibinfo{author}{\bibfnamefont{K.}~\bibnamefont{Pachucki}},
  \bibinfo{journal}{Phys. Rev. A} \textbf{\bibinfo{volume}{78}},
  \bibinfo{pages}{012504} (\bibinfo{year}{2008}).

\bibitem[{\citenamefont{Grotch}(1970)}]{Grotch1970}
\bibinfo{author}{\bibfnamefont{H.}~\bibnamefont{Grotch}},
  \bibinfo{journal}{Phys. Rev. Lett.} \textbf{\bibinfo{volume}{24}},
  \bibinfo{pages}{39} (\bibinfo{year}{1970}).

\bibitem[{\citenamefont{Czarnecki et~al.}(1996)\citenamefont{Czarnecki, Krause,
  and Marciano}}]{Czarnecki96}
\bibinfo{author}{\bibfnamefont{A.}~\bibnamefont{Czarnecki}},
  \bibinfo{author}{\bibfnamefont{B.}~\bibnamefont{Krause}}, \bibnamefont{and}
  \bibinfo{author}{\bibfnamefont{W.~J.} \bibnamefont{Marciano}},
  \bibinfo{journal}{Phys. Rev. Lett.} \textbf{\bibinfo{volume}{76}},
  \bibinfo{pages}{3267} (\bibinfo{year}{1996}).

\bibitem[{\citenamefont{Prades et~al.}(2010)\citenamefont{Prades, de~Rafael,
  and Vainshtein}}]{Prades10}
\bibinfo{author}{\bibfnamefont{J.}~\bibnamefont{Prades}},
  \bibinfo{author}{\bibfnamefont{E.}~\bibnamefont{de~Rafael}},
  \bibnamefont{and}
  \bibinfo{author}{\bibfnamefont{A.}~\bibnamefont{Vainshtein}},
  \emph{\bibinfo{title}{The hadronic light-by-light scattering contribution to
  the muon and electron anomalous magnetic moments}} (\bibinfo{publisher}{World
  Scientific}, \bibinfo{address}{Singapore}, \bibinfo{year}{2010}),
  vol.~\bibinfo{volume}{20} of \emph{\bibinfo{series}{Advanced Series on
  Directions in High Energy Physics}}, chap.~\bibinfo{chapter}{9}, pp.
  \bibinfo{pages}{303--317}.

\bibitem[{\citenamefont{Nomura and Teubner}(2013)}]{Nomura13}
\bibinfo{author}{\bibfnamefont{D.}~\bibnamefont{Nomura}} \bibnamefont{and}
  \bibinfo{author}{\bibfnamefont{T.}~\bibnamefont{Teubner}},
  \bibinfo{journal}{Nucl. Phys. B} \textbf{\bibinfo{volume}{867}},
  \bibinfo{pages}{236} (\bibinfo{year}{2013}).

\bibitem[{\citenamefont{Kurz et~al.}(2014)\citenamefont{Kurz, Liu, Marquard,
  and Steinhauser}}]{Kurz14}
\bibinfo{author}{\bibfnamefont{A.}~\bibnamefont{Kurz}},
  \bibinfo{author}{\bibfnamefont{T.}~\bibnamefont{Liu}},
  \bibinfo{author}{\bibfnamefont{P.}~\bibnamefont{Marquard}}, \bibnamefont{and}
  \bibinfo{author}{\bibfnamefont{M.}~\bibnamefont{Steinhauser}},
  \bibinfo{journal}{Phys. Lett. B} \textbf{\bibinfo{volume}{734}},
  \bibinfo{pages}{144} (\bibinfo{year}{2014}).

\bibitem[{\citenamefont{Furry}(1951)}]{Furry1951}
\bibinfo{author}{\bibfnamefont{W.~H.} \bibnamefont{Furry}},
  \bibinfo{journal}{Phys. Rev.} \textbf{\bibinfo{volume}{81}},
  \bibinfo{pages}{115} (\bibinfo{year}{1951}).

\bibitem[{\citenamefont{Pohl et~al.}(2017)\citenamefont{Pohl, Nez, Fernandes,
  Ahmed, Amaro, Amaro, Biraben, Cardoso, Covita, Dax et~al.}}]{Pohl2017}
\bibinfo{author}{\bibfnamefont{R.}~\bibnamefont{Pohl}},
  \bibinfo{author}{\bibfnamefont{F.}~\bibnamefont{Nez}},
  \bibinfo{author}{\bibfnamefont{L.~M.~P.} \bibnamefont{Fernandes}},
  \bibinfo{author}{\bibfnamefont{M.~A.} \bibnamefont{Ahmed}},
  \bibinfo{author}{\bibfnamefont{F.~D.} \bibnamefont{Amaro}},
  \bibinfo{author}{\bibfnamefont{P.}~\bibnamefont{Amaro}},
  \bibinfo{author}{\bibfnamefont{F.}~\bibnamefont{Biraben}},
  \bibinfo{author}{\bibfnamefont{J.~M.~R.} \bibnamefont{Cardoso}},
  \bibinfo{author}{\bibfnamefont{D.~S.} \bibnamefont{Covita}},
  \bibinfo{author}{\bibfnamefont{A.}~\bibnamefont{Dax}}, \bibnamefont{et~al.},
  in \emph{\bibinfo{booktitle}{Proceedings of the 12th International Conference
  on Low Energy Antiproton Physics (LEAP2016)}} (\bibinfo{year}{2017}), p.
  \bibinfo{pages}{011021}.

\bibitem[{\citenamefont{Sturm et~al.}(2017)\citenamefont{Sturm, Vogel,
  K\"ohler-Langes, Quint, Blaum, and Werth}}]{Sturm2017}
\bibinfo{author}{\bibfnamefont{S.}~\bibnamefont{Sturm}},
  \bibinfo{author}{\bibfnamefont{M.}~\bibnamefont{Vogel}},
  \bibinfo{author}{\bibfnamefont{F.}~\bibnamefont{K\"ohler-Langes}},
  \bibinfo{author}{\bibfnamefont{W.}~\bibnamefont{Quint}},
  \bibinfo{author}{\bibfnamefont{K.}~\bibnamefont{Blaum}}, \bibnamefont{and}
  \bibinfo{author}{\bibfnamefont{G.}~\bibnamefont{Werth}},
  \bibinfo{journal}{Atoms} \textbf{\bibinfo{volume}{5}} (\bibinfo{year}{2017}).

\end{thebibliography}
\end{document}